\documentclass[12pt]{amsart}
\usepackage{amsmath}
\usepackage{amsthm}
\usepackage{amsfonts}
\usepackage{graphicx}
\newenvironment{nouppercase}{
  
  \renewcommand{\uppercasenonmath}[1]{}}{}

\begin{document}

\title
{Composite dynamical symmetry of M--branes}
\author{Jens Hoppe}
\address{Braunschweig University, Germany}
\email{jens.r.hoppe@gmail.com}

\begin{abstract}
It is shown that the previously noticed internal dynamical $SO(D-1)$ symmetry \cite{1} for relativistic M--branes moving in $D$--dimensional space--time is naturally realized in the (extended by powers of $\frac{1}{p_+}$) enveloping algebra of the Poincar\'e algebra.
\end{abstract}

\begin{nouppercase}
\maketitle
\end{nouppercase}
\thispagestyle{empty}
\noindent
In the common light--cone derivation of the critical dimension for bosonic strings it is hidden in the calculation that the identification of terms in $M_{i-}$ {\it not} involving zero--modes does not only require subtracting $X_i P_- - P_i X_-$ but also terms that are linear (!) in the transverse total momenta, implicit in the longitudinal oscillators (see e.g. \cite{GGRT}\cite{GA}). The purely internal parts of ($P_+$times) the longitudinal Lorentz--generators  $M_{i-}$, and the $M_{ij}$ (generators of $SO(D-2)$), satisfy     
\begin{equation}\label{eq1} 
\begin{split}
\lbrace \mathbb{M}_{jk}, \mathbb{M}_{i-}\rbrace & = -\delta_{ik}\mathbb{M}_{j-} + \delta_{ij}\mathbb{M}_{k-}\\
\lbrace \mathbb{M}_{i-}, \mathbb{M}_{j-}\rbrace & = \mathbb{M}^2 \cdot \mathbb{M}_{ij}\\
\lbrace \mathbb{M}_{ij}, \mathbb{M}_{kl}\rbrace & = -\delta_{jk}\mathbb{M}_{il} \pm 3 \, \text{more}
\end{split}
\end{equation}
with $\mathbb{M}^2 = 2P_+P_- - \vec{P}^2$ the internal $(\text{Mass})^2$, very similar to the dynamical symmetry of the hydrogen atom -- which gives hope \cite{1} that it may be possible to obtain purely algebraically the spectrum of $\mathbb{M}^2$ (when quantized),
with the dimension and the topology of the extended object being encoded in the dimensions and multiplicities of the occurring irreducible finite dimensional representations of $SO(D-1)$ given by (\ref{eq1}) via $\mathbb{L}_{i-} := \frac{\mathbb{M}_{i-}}{\sqrt{\mathbb{M}^2}}$ and $\mathbb{L}_{ij} := \mathbb{M}_{ij}$.\\
Attempts to quantize (\ref{eq1}), using the constrained phase--space of transverse internal degrees of freedom are hindered by the constraints (that are reflecting residual invariance of the theory under volume--preserving diffeomorphismus, resp. solvability for the longitudinal degrees of freedom in terms of the transverse ones) -- making even classical calculations, like the proof \cite{G} of Poisson commutativity of the $M_{i-}$ formidable.
In \cite{BH}, on the other hand, it was noticed that in the codimension one case (to which we intermediately restrict) relativistic M(em)--branes can be described as an isentropic inviscid irrotational gas. Taking proper care of (cp.\cite{H2}) 
\begin{equation}\label{eq2} 
P_+ := \int \sqrt{\frac{g}{2\dot{\zeta}-\dot{\vec{x}}\,^2}}\,d^M\varphi = \eta \int \rho\, d^M\varphi = \eta = \int q\,d^M x
\end{equation}
when performing the hodograph--transformation
\begin{equation}\label{eq3} 
\begin{split}
\varphi^{\alpha} & = (\tau, \varphi^1, \ldots , \varphi^M) \rightarrow x^{\alpha} = (\tau, x^1(\tau, \varphi ), \ldots , x^M(\tau, \varphi ))\\
\vert \frac{\partial x^{\alpha}}{\partial \varphi ^{\beta}} \vert & = \vert \frac{\partial x^i}{\partial \varphi ^b} \vert  = \rho \lbrace x_1 , \ldots, x_M \rbrace =: \frac{\eta \rho}{q(\vec{x}, \tau)}\\
 1 & = \int \rho\,d^M\varphi = \int \frac{\rho}{\vert \frac{\partial x}{\partial \varphi}\vert}\,d^M x = \frac{1}{\eta}\int q\,d^M x\\
\int f(\varphi^{\alpha}) \rho \,d^M \varphi & = \frac{1}{\eta}\int \hat{f}(x^{\alpha})q\,d^M x\\
\frac{\vec{p}}{\eta\rho}(\varphi^{\alpha}) & = (\vec{\nabla}p)_{(x(\varphi))},\\
X_- & = \zeta_0 = \int \zeta \rho\, d^M\varphi \stackrel{!}{=} \frac{1}{\eta}\int pq\,d^M x = -\frac{L_{+-}}{P_+} + \tau \frac{P_-}{P_+} \\
X_i & =\int x_i\rho\,d^M \varphi = \frac{1}{\eta} \int x_i q = \frac{L_{i+}}{P_+} + \tau \frac{P_i}{P_+}\\
P_- & = \frac{1}{2\eta} \int \big(\frac{\vec{P}^2}{\rho^2} + \lbrace \, , \ldots , \, \rbrace^2 \big)\rho\,d^M \varphi = \frac{1}{2}\int \big((\vec{\nabla}p)^2 + \frac{1}{q^2}\big)q\,d^M x
\end{split}
\end{equation}
one obtains the hydrodynamic M--brane Poincar\'e--generators (\cite{BH}\cite{AJ}\cite{H2})
\begin{equation}\label{eq4} 
\begin{split}
P_- & = \frac{1}{2} \int \big( q(\nabla p)^2 + \frac{1}{q} \big),\quad P_+ = \int q, \quad \vec{P} = \int q\vec{\nabla p}\\
L_{ab} & = \int q (x_a \partial_b p - x_b \partial_a p), \quad L_{a+} = \int q x_a - \tau P_a\\
L_{a-} & = \frac{1}{2} \int \big(x_a(q (\nabla p)^2 + \frac{1}{q}) - q \partial_a(p^2)\big)\\
L_{+-} & = -\int q p\, d^M x + \tau P_- 
\end{split}
\end{equation}
satisfying
\begin{equation}\label{eq5} 
\begin{split}
\lbrace L_{a\pm}, L_{+-}\rbrace & = \mp L_{a\pm}, \quad \lbrace L_{a-}, L_{b-}\rbrace = 0\\
\lbrace L_{a\pm}, P_{\mp}\rbrace & = P_a, \quad \lbrace L_{+-}, P_{\pm}\rbrace = \pm P_{\pm}\\
\lbrace L_{ab}, L_{cd}\rbrace & = -\delta_{bc} L_{ad} \pm 3 \, \text{more}\\  
\lbrace L_{a+}, L_{b-}\rbrace & = \delta_{ab}L_{+-} - L_{ab}.
\end{split}
\end{equation}
Due to the zero--modes being ratios\footnote{up to terms proportional to $\tau$ (which drop out in (\ref{eq7}))} of $SO(D-1, 1)$ generators, 
\begin{equation}\label{eq6} 
\begin{split}
X_i & = \frac{L_{i+}}{P_+} + \tau \frac{P_i}{P_+},\\
X_- & = \zeta_0 = -\frac{L_{+-}}{P_+} + \tau \frac{P_-}{P_+}
\end{split}
\end{equation}
one may write the internal (`$SO(D-1)$') generators occurring in (\ref{eq1}) as composite operators, solely as rational\footnote{were it not for the $\frac{1}{P_+}$ in the subtraction to $M_{ij}$, and eventual appearance of (commuting) factors of $\frac{1}{\sqrt{\mathbb{M}^2}}$, {\it polynomial}, i.e. elements of the enveloping algebra} expression in the generators of the original Poincar\'e algebra:
\begin{equation}\label{eq7} 
\begin{split}
\mathbb{M}_{ij} & = L_{ij} - \frac{1}{P_+}(L_{i+}P_j - L_{j+}P_i) = L_{ij} - L'_{ij}\\
\mathbb{M}_{i-} & = P_+L_{i-} - (\underbrace{L_{i+}P_- + L_{+-}P_i}_{=P_+ L'_{i-}} + \underbrace{\mathbb{M}_{ik}P_{k}}_{=P_+\tilde{L}'_{i-}})\\
& =: P_+L_{i-} - P_+\tilde{L}_{i-};
\end{split}
\end{equation}
and a tedious, but straightforward calculation, resp. 
\begin{equation}\label{eq8} 
\begin{split}
\lbrace L'_{ij}, L'_{kl}\rbrace & = -\delta_{jk} L'_{il} \pm 3 \, \text{more},\quad 
\lbrace L_{ij}, L'_{kl}\rbrace  = -\delta_{jk} L'_{il} \pm 3 \, \text{more},\\ 
\lbrace L'_{i-}, L'_{j-}\rbrace & = 0,\\
\lbrace \tilde{L}'_{i-},  \tilde{L}'_{j-}\rbrace & = \frac{1}{P^2_+}(\vec{P}^2\mathbb{M}_{ij} + P_i\mathbb{M}_{jk}P_k - P_j\mathbb{M}_{il}P_l)\\
\lbrace L'_{i-}, \tilde{L}'_{j-}\rbrace & - (i \leftrightarrow j) = -2\frac{P_-}{P_+}\mathbb{M}_{ij} - \frac{1}{P^2_+} (P_i\mathbb{M}_{jk}P_k  - P_j\mathbb{M}_{il}P_l)\\
\lbrace \tilde{L}_{i-},  \tilde{L}_{j-}\rbrace & = -\frac{\mathbb{M}^2}{P^2_+}\mathbb{M}_{ij},
\quad \lbrace \frac{\mathbb{M}_{i-}}{P_+}, \tilde{L}_{j-} \rbrace  - (i \leftrightarrow j)  = 0,
\end{split}
\end{equation}
 note also 
\begin{equation}\label{eq9} 
\lbrace \mathbb{M}_{ik}, P_j \rbrace  = 0,\quad
\lbrace \mathbb{M}_{ik}, L_{+-} \rbrace = 0, \quad
\lbrace P_+, \mathbb{M}_{i-} \rbrace = 0,
\end{equation}
gives that (\ref{eq7}) indeed satisfies (\ref{eq1}).

\vspace{1.5cm}
\noindent
\textbf{Acknowledgement.} I would like to thank A. Jevicki for discussions.

\end{document}